\documentclass[runningheads]{llncs}
\usepackage{graphicx}

\usepackage{longtable}
\usepackage{booktabs}
\usepackage{bm}
\usepackage{color}
\usepackage{amsmath}
\usepackage{amssymb}
\usepackage{gensymb}

\begin{document}
\title{Segmentation for Classification of Screening Pancreatic Neuroendocrine Tumors}
\titlerunning{Segmentation for Classification of Screening PNETs}
\authorrunning{Z. Zhu \emph{et al.}}
\author{Zhuotun Zhu\textsuperscript{1}, Yongyi Lu\textsuperscript{1}, \\
Wei Shen\textsuperscript{1}, Elliot K. Fishman\textsuperscript{2}, Alan L. Yuille\textsuperscript{1}}

\institute{
\textsuperscript{1}The Johns Hopkins University, Baltimore, MD 21218, USA\\
{\tt\small \{zhuotun,yylu1989,shenwei1231,alan.l.yuille\}@gmail.com}\\
\textsuperscript{2}The Johns Hopkins University School of Medicine, Baltimore, MD 21287, USA\\
{\tt\small efishman@jhmi.edu}\\
}

\maketitle              
\begin{abstract}
This work presents comprehensive results to detect in the early stage the pancreatic neuroendocrine tumors (PNETs), a group of endocrine tumors arising in the pancreas, which are the second common type of pancreatic cancer, by checking the abdominal CT scans. To the best of our knowledge, this task has not been studied before as a computational task. To provide radiologists with tumor locations, we adopt a segmentation framework to classify CT volumes by checking if at least a sufficient number of voxels is segmented as tumors. To quantitatively analyze our method, we collect and voxelwisely label a new abdominal CT dataset containing $376$ cases with both arterial and venous phases available for each case, in which $228$ cases were diagnosed with PNETs while the remaining $148$ cases are normal, which is currently the largest dataset for PNETs to the best of our knowledge. In order to incorporate rich knowledge of radiologists to our framework, we annotate dilated pancreatic duct as well, which is regarded as the sign of high risk for pancreatic cancer. Quantitatively, our approach outperforms state-of-the-art segmentation networks and achieves a sensitivity of $89.47\%$ at a specificity of $81.08\%$, which indicates a potential direction to achieve a clinical impact related to cancer diagnosis by earlier tumor detection.

\end{abstract}

\section{Introduction}
The American Cancer Society estimates that about $56\rm{,}770$ people in the United States will be diagnosed with pancreatic cancer in 2019, and that $45\rm{,}750$ will die from the disease~\cite{acs2019cancer}. The oncology community has expended arsenal at this disease with little effect: the $5$-year survival rate remains at only $\approx5\%$~\cite{stewart2017world} despite decades of effort. This is due to the fact that most patients with localized disease have no recognizable symptoms or signs; as a result, upon diagnosis, tumors have generally spread to critical abdominal vessels and/or adjacent organs, which is too late to be cured. Despite the grim statistics, there is still real hope for the early detection, which can boost the $5$-year survival rate by $3$ times to reach around $20\%$ given an early diagnosis~\cite{board2017pancreatic}. Among pancreatic cancers, the pancreatic adenocarcinoma (PDAC) is the most common type of pancreatic cancer. Recently, there is a study~\cite{zhu2018multi} showing that they can achieve an overall sensitivity of $94.1\%$ at a specificity of $98.5\%$ for the detection of PDAC, which sheds light on the possibility of early pancreatic cancer detection. In our work, we focus on the early detection of pancreatic neuroendocrine tumors (PNETs) from CT scans, which is harder than the detection of PDAC considering PNETs are less common with even smaller voxel size.

The detection of PNETs falls into the area of computer aided diagnosis. The main challenges are lying in three folds: $1$) the small size of tumors with respect to the whole volume; $2$) the large tumor variations in location, shape and size across different patients; $3$) the abnormalities can change the texture of surrounding tissues a lot, which makes the task even harder to locate the tumor targets. With the unprecedented development of deep learning, in particularly fully convolutional neural networks (FCNs), there are works which has been driving the field forward in image segmentation~\cite{chen2016deeplab,milletari2016v,ronneberger2015u}. In the pancreas segmentation area, researchers have been actively pushing the boundaries of obtaining accurate segmentation performance on both normal pancreas~\cite{cai2017improving,roth2016spatial} and abnormal pancreas~\cite{zhou2017deep,zhu2018a}. 

Valuable insights from radiologists' clinical diagnosis and analysis process can be leveraged to tackle this problem. {\bf{First}}, radiologists are very sensitive to the dilated pancreatic duct when reading CT scans. There are often occasions the pancreatic duct is visible to be dilated though the PNETs are barely visible from CT appearance and texture, in which case the PNETs are correctly imagined to be at the cut-off region of the dilated pancreatic duct. {\bf{Second}}, the appearance and texture cues can be very different for PNETs between the venous and arterial phases of CT scans. Radiologists make diagnosis decision by checking both phases in case the PNETs are hardly to be picked up in one phase. Missing true PNETs can cause critical areas to remain untreated. To migrate the aforementioned practical knowledge from radiologists routine works to our system, on the one hand, we annotated the voxels of dilated pancreatic duct as strong auxiliary cues to indicate the present of pancreatic cancer. On the other hand, we conduct PNETs segmentation and classification on both arterial and venous phases to reduce the missing detection of PNETs. This is done quite different from the state-of-the-art work on PDACs~\cite{zhu2018multi}, where they only study on one phase and no cues are explored from the dilated pancreatic duct. Our final goal is aimed to detect PNETs from a mixed set of normal and abnormal CT scans. It is not a simple binary classification task because radiologists want to know the location of tumors, so we use the idea of Segmentation-for-Classification (S4C), which trains segmentation models and uses voxel predictions for the classification.

Our {\bf contributions} are three folds: $1$) we voxelwisely label a dual-phase PNETs dataset in both arterial and venous phases, which is the largest dataset and study up-to-date to the best of our knowledge; $2$) we are the first work of segmentation and classification for PNETs, where the segmentation makes the classification task interpretable and extra cues from the dilated pancreatic duct are incorporated in proposed framework; $3$) our overall framework achieves a sensitivity of $89.47\%$ at a specificity of $81.08\%$, which indicates the potential direction to a clinical impact.

\section{Method}
\subsection{The Overall Framework}

We denote the dataset as ${\mathbf{D}}={\left\{\left(\mathbf{X}_1,y_1\right),\ldots,\left(\mathbf{X}_N,y_N\right)\right\}}$, where $N$ is total number of CT cases, ${\mathbf{X}_n}\in{\mathbb{R}^{W_n\times H_n\times L_n}}$ is a $3$D volume with each voxel defined as the Hounsfield Unit (HU), and ${y_n}\in{\left\{0,1\right\}}$ is the case label, by which $0$ means a normal case while $1$ for an abnormal case. By abnormal/tumor we mean cases diagnosed with PNETs throughout the whole paper. Our goal is to design a model $\mathbb{M}:{y}={f\!\left(\mathbf{X}\right)}$ mapping a CT image to its state of being abnormal or not. 

Some previous work suggested to classify medical images by directly using deep neural networks~\cite{dou2017multilevel,hussein2018supervised}, however, we claim that a better strategy is to perform tumor segmentation together with the classification task. This makes the prediction {\bf interpretable} for the classification results from segmentation cues, by which radiologists can take a further investigation of the suspicious abnormal regions. But for a deep neural network doing direct classification, it is hard for radiologists to further check which regions are suspicious while the adopted segmentation-for-classification (S4C)~\cite{zhu2018multi} sheds light on the abnormality detection, which is more plausible. In addition, this harnesses voxelwise annotations as fully supervision into the classification model, so that the entire network can be better optimized. Different from~\cite{zhu2018multi} which did S4C for PDAC only on venous phase, we incorporate the dilated pancreatic duct information on both arterial and venous phases for PNETs, which can further improve the sensitivity.

On our dataset, each training case is associated with a segmentation mask $\mathbf{M}_n$ of the same dimension as $\mathbf{X_n}$, among which ${m_{n,i}}\in{\left\{0,1,2,3\right\}}$ denotes the annotated categories for the $i$-th voxel. More specifically, a background voxel is labelled as $0$, and $1$ means the voxel is inside the normal pancreas regions, and $2$ denotes a voxel inside the tumor regions. We would like to emphasize that besides normal/abnormal pancreas regions and background voxels, we annotate voxels inside dilated pancreatic duct regions as $3$. This is motivated from the knowledge of radiologists that a pancreatic duct dilation is a sign of high risk for pancreatic cancer. If we predict a dilated pancreatic duct present for some cases where the PNETs are hardly invisible from textures, we can refine our judgment and would not miss those really hard cases. Note that a pancreas set includes the normal pancreas set, abnormal pancreas set and the dilated pancreatic duct set. The segmentation module is a mapping function ${\mathbf{M}}={\mathbf{s}\!\left(\mathbf{X}\right)}$, which is implemented by an encoder-decoder network mapping from CT scans with Hounsfield scale values to the categorical sets. The classification module is a binary function ${y}={c\!\left(\mathbf{M}\right)}$ with a set of rules given the segmentation as input that we will elaborate later. All in all, the whole framework is denoted as: 
\begin{equation}
\label{Eqn:Framework}
{y}={f\!\left(\mathbf{X}\right)}={c\circ\mathbf{s}\!\left(\mathbf{X}\right)}.
\end{equation}

\begin{figure}[t]
	\begin{center}
		\includegraphics[width=1.0\linewidth]{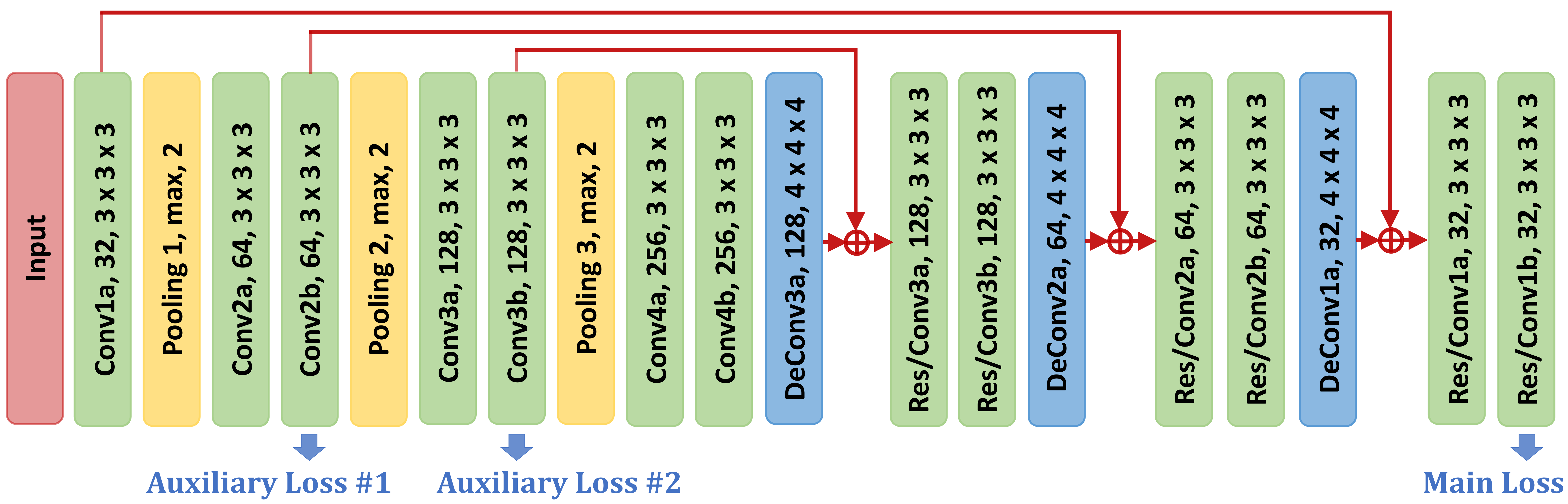} \\
	\end{center}
	\caption{The network backbone of our S4C pipeline. We adopt an encoder-decoder fashion, where the encoder path on the left acts as a feature extractor to learn more and more compact features while the decoder path on the right decompresses the learned features gradually to obtain the dense predictions with higher and higher resolutions. The sum residual connections from the low-level layers are crucial to integrate the pixel-level features such as edges to the semantically meaningful features of high-level layers such as patterns or shapes. The two auxiliary losses serve as a deep supervision to reach a better optimization process, which favors the final segmentation performance~\cite{zhu2018a}. The whole network is optimized with voxelwise softmax cross-entropy loss. The weight ratio for auxiliary losses \#1, \#2 and the main loss is $1:2:5$. Best viewed in color.}
	\label{Fig:ResDSN}
\end{figure}

\vspace{-0.5cm}
\subsection{Segmentation for Classification (S4C)}
Our segmentation backbone is shown in Fig.~\ref{Fig:ResDSN}, which adopts the encoder and decoder~\cite{ronneberger2015u} fashion for the dense prediction. The residual connections and auxiliary losses are the delicate designs aimed at a better and stable optimization~\cite{zhu2018a}. The pooling layers of the encoder path compress the learning process into more compact feature space, from where the DeConv layers of the decoder path decompress them to semantically meaningful features in the fine-scale resolution. The whole framework takes the voxelwise softmax cross-entropy as the loss function, which shows stable and supreme performance on both normal pancreas and cystic pancreas segmentation~\cite{zhu2018a}. The segmentation network takes patches as input, whose size is set to be $64$$\times$$64$$\times$$64$, which covers sufficient context and makes memory for the networks design with powerful representation ability.

During training, we implemented simple yet effective augmentations on patches input, \emph{i.e.}, rotation ($90\degree, 180\degree, \textrm{ and } 270\degree$) and flip in all three axes (axial, sagittal and coronal), to increase the number of training samples which can alleviate the limited number of CT cases with annotations. During testing, we adopted the sliding window way to slide the whole CT volumes with a $20$-voxel spatial stride. The overlapped regions are voted by majority. Based on the segmentation prediction, we classify each volume to be abnormal or normal. We compute the maximal connected component $\mathcal{C}_{\textrm{max}}$ and keep a component which is either larger than $20\%$ of $\mathcal{C}_{\textrm{max}}$ or at a distance of less than $27$ voxels to $\mathcal{C}_{\textrm{max}}$. As for classification, a volume is predicted as PNETs if as least $40$ voxels are predicted as tumors or $500$ voxels are predicted as dilated pancreatic duct. To harness the dual-phase information, we classify a CT case as abnormal given any phase is predicted as PNETs, which improves the sensitivity at the cost of the specificity.

\begin{figure}[t]
	\begin{center}
		\includegraphics[width=1.0\linewidth]{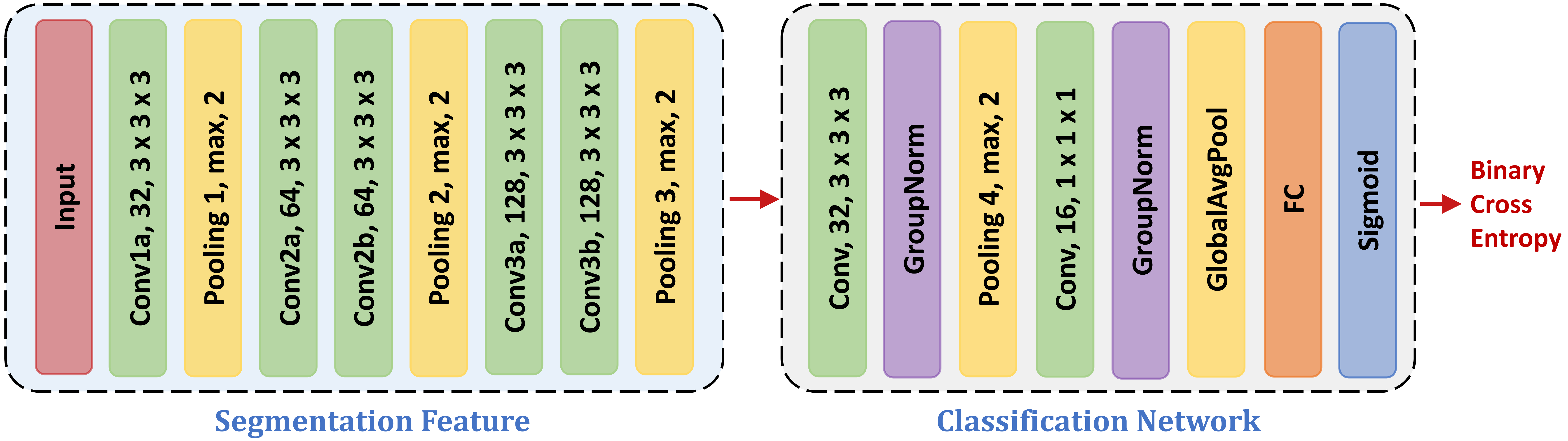} \\
	\end{center}
	\caption{The classification network designed for the direct binary classification, \emph{i.e.}, tumor versus non-tumor, as an ablation study. Best viewed in color.}
	\label{Fig:Classifier}
\end{figure}

\subsection{Classification Network as Comparison}
Since we adopt the Segmentation-for-Classification framework as our pipeline, it is natural to see how a classification network performs in comparison. Therefore, we implement a classification network as shown in Fig~\ref{Fig:Classifier}. To compare the S4C and classification network as fair as possible, we construct the classification network by feeding the features maps of segmentation network as input. This is due to the fact that the classification label is $0/1$ per CT case, which owns much less information than the voxelwise $0/1$'s of segmentation labels. To filter out the large out-of-pancreas regions during training the classification network, an $128$-way feature vector is extracted from the Region-of-Interest (RoI) of pancreas, which is derived from the ground-truth in the training or from the segmentation prediction with a margin in the testing. The $128$-dimension Pool3 feature is chosen rather than the $256$-dimension Conv4b feature vectors because of the better generalization ability we observed during our experiments. Since the feature map size is different for different size of pancreas RoIs, the batch size is chosen to be $1$, then a GroupNorm~\cite{wu2018group} is added after each convolution in the classification to help the learning process. Note that the segmentation of pancreas is very good, which makes it doable for the RoI as input.

\section{Experiment}
\subsection{Implementation Details}
We collected a new dataset with $376$ cases in total from potential renal donors, in which we have $148$ normal cases and $228$ biopsy-proven PNETs cases, where each case has both arterial and venous phases available. Four experts in abdominal anatomy voxelwisely annotated the pancreas, tumor regions, and dilated pancreatic duct using the Varian Velocity software, and checked by an experienced board-certified abdominal radiologist. For a radiologist expert, an average normal case took $20$ minutes, and an average abnormal case $40$ minutes to finish the voxelwise annotation. To quantitatively analyze our method, we adopt a same $4$-fold cross-validation for S4C and classification network on these $376$ cases in both phases. All in all, for a single phase, each training set contains $111$ normal and $171$ abnormal cases, and each corresponding testing set contains $57$ abnormal and $37$ normal cases. And the final quantitative performance is reported on the testing of all cases across $4$ folds, by which we take every case into consideration to fully maximize the utilization of the medical data which are expensive and time-consuming to obtain. Our framework is implemented on Pytorch $0.5.0$, and the GPU we are running on is the Tesla $\text{V}100$. The base learning rate is $0.01$ and decayed polynomially (the power is $0.9$) in a total of $80\rm{,}000$ iterations with a batch size of $16$ for the S4C. The base learning rate is $0.001$ and decayed polynomially (the power is $0.9$) in a total of $20\rm{,}000$ iterations for the classification network. The weight decay and momentum are set to be $0.0005$ and $0.9$, separately. The total training time for a S4C model is $2.5$ days while the average testing time for a case is around $10$ mins while the training time for a classification is $\approx12$ mins given the segmentation features as input. All parameters are verified by the $4$-fold cross-validation.

One of our goals is to quantify the segmentation accuracy by the Dice-S{\o}rensen Coefficient (DSC) between the predicted and the ground-truth tumor regions $\mathcal{Y}$ and $\mathcal{Y}^\star$, {\em i.e.}, ${\mathrm{DSC}\!\left(\mathcal{Y},\mathcal{Y}^\star\right)}={\frac{2\times\left|\mathcal{Y}\cap\mathcal{Y}^\star\right|}{\left|\mathcal{Y}\right|+\left|\mathcal{Y}^\star\right|}}$. Our primary goal is to measure the abnormality classification by the sensitivity (the percentage of correctly classified abnormal cases) and the specificity (the percentage of correctly classified normal cases). In practice, there is always a trade-off between the sensitivity and specificity. We care much more about the sensitivity than the specificity since the final goal is to detect PNETs in the early stage for timely medical interventions.

\subsection{Performance}
\begin{figure}[t]
	\begin{center}
		\includegraphics[width=1.0\linewidth]{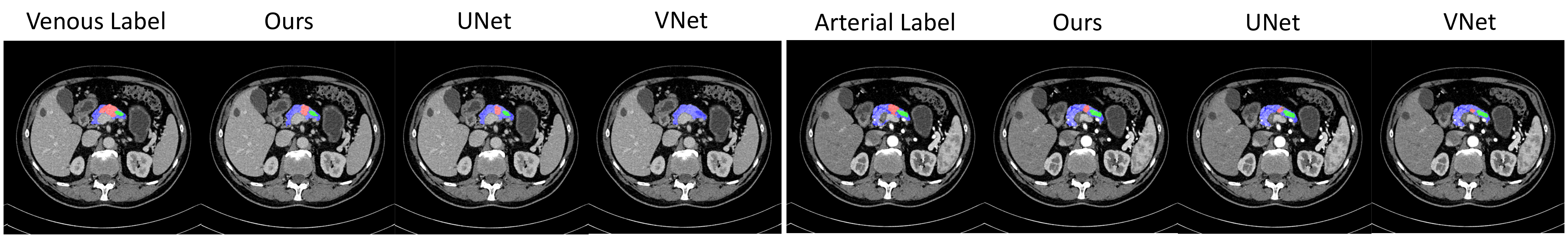} \\
	\end{center}
	\caption{The segmentation visualization for the case number 7263. ``Ours" method successfully detects the PNETs and dilated pancreatic duct regions on both the venous and the arterial phase, which performs better then ``3D UNet" and ``VNet".}
	\label{Fig:7263Main}
\end{figure}

From Table~\ref{Tab:Results} that shows single-phase results, considering venous and arterial phases equally, our method in the venous phase achieves the best results on all evaluation matrix except for the comparable result with 3D UNet on the venous normal pancreas segmentation. On the one hand, the pancreas segmentation can be as high as $87.41\%$ and $84.69\%$ for the normal pancreas and abnormal pancreas respectively, which demonstrates the effectiveness of our method. On the other hand, the tumor segmentation performance is promising to be $43.11\%$, which outperforms the state-of-the-art segmentation networks, \emph{i.e.}, 3D UNet~\cite{cciccek20163d} and VNet~\cite{milletari2016v}. From the tumor segmentation results, it is a really challenging segmentation problem considering the various size, shape and locations of tumors. Note that a recent work on the PDAC segmentation achieves a DSC of $56.46\pm{26.23}\%$~\cite{zhu2018multi}, which is not as hard as the PNETs which are less common with even smaller voxel size. As for the abnormality classification task, our single phase model achieves $82.46\%$ sensitivity at a specificity of $91.89\%$, which beats the second best 3D UNet by $0.88\%$ and $2.70\%$ respectively. To compare the same model in arterial and venous phases, we find that all three models behave generally better on the venous phase than the arterial phase. As in Fig.~\ref{Fig:7263Main}, our method performs better segmentation results for both venous and arterial phases, which shows the more powerful representation ability of our network backbone.

\begin{figure}[t]
	\begin{center}
		\includegraphics[width=0.8\linewidth]{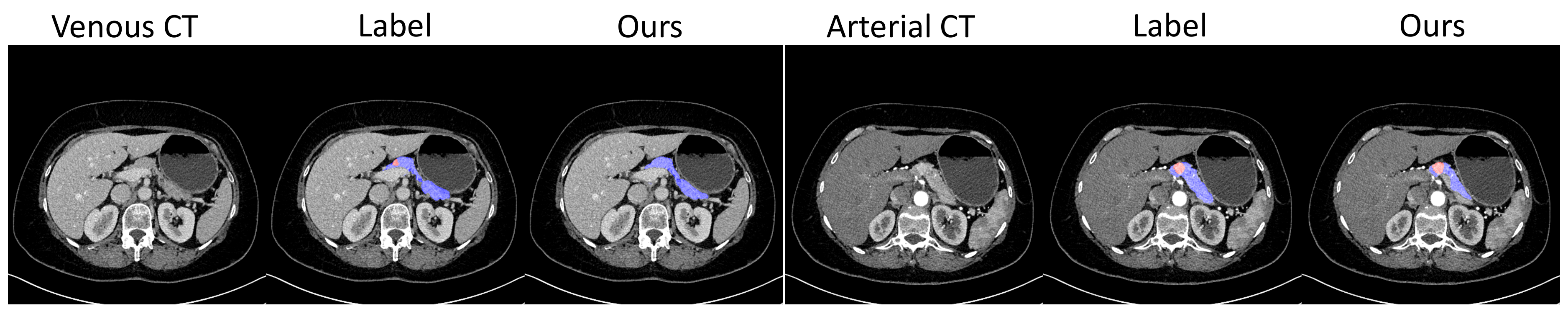} \\
	\end{center}
	\caption{The segmentation visualization for the case number 7264. The tiny PNETs is hanging of the pancreas head, where ``Ours" method successfully detects the PNETs regions on the arterial phase while missing the detection on the venous phase.}
	\label{Fig:7264Main}
\end{figure}

\begin{table}[t]

 \begin{center}
 \begin{tabular}{ccccccccc}\toprule
      Phase &Method&N.Pancreas  &A.Pancreas & Tumor &Misses &W.Calls& Sensitivity & Specificity \\
\hline
   Venous &Ours &$87.41\%$ &$\mathbf{84.69\%}$ &$\mathbf{43.11\%}$  &$\mathbf{40/228}$ &$\mathbf{12/148}$ & $\mathbf{82.46\%}$ &   $\mathbf{91.89\%}$ \\
   Venous &3D UNet &$\mathbf{87.70\%}$ &$83.84\%$ &$41.43\%$  &$42/228$ &$16/148$ & $81.58\%$ &   $89.19\%$ \\
   Venous &VNet &$86.76\%$ &$83.90\%$ &$40.67\%$  &$49/228$ &$20/148$ & $78.51\%$ &   $86.49\%$ \\
\hline

   Arterial &Ours &$81.78\%$ &$83.34\%$ &$42.49\%$ &$44/228$ &$21/148$ & $80.70\%$ & $85.81\%$\\

   Arterial &3D UNet &$82.47\%$ &$83.33\%$ &$42.58\%$ &$44/228$ &$20/148$ & $80.70\%$ & $86.49\%$\\
   Arterial &VNet &$83.85\%$ &$82.79\%$ &$39.22\%$ &$43/228$ &$31/148$ & $81.14\%$ & $79.05\%$\\
 \bottomrule
 \end{tabular}
 \end{center}
 \caption{Performance of segmentation and classification on our own dataset in two different phases. From left to right: normal pancreas cases, abnormal pancreas cases and tumor segmentation accuracy (DSC, $\%$), the number of missed abnormal cases out of $228$ abnormal cases in total, the number of wrong calls of tumor predictions out of $148$ normal cases in total, the corresponding sensitivity and the specificity.}
 \label{Tab:Results}
\end{table}

\subsection{Dual-Phase Fusion and Comparison with Classification Network}

In the clinical practice, the radiologists generally care much more about the sensitivity than they do about the specificity when it comes to the tumor detection. In radiology, some tissues are more visible in the venous phase while others are better to be captured in the arterial phase. Given we have CT scans in both arterial and venous phases available for each case, it is natural to think that we can combine the detection results from two phases together to take advantage of the different enhancement patterns when detecting the abnormality from different phases. We come up with a very straightforward way to combine the detection results. More specifically, if a model trained on any phase predicts this case to be abnormal, we regard this case to be abnormal. In this way, we are able to reduce the missing cases since a PNETs case can only be missed if both two phases classify the same case to be normal. The quantitative results are given in Table~\ref{Tab:FusePhaseResults}. First, our model beats both 3D UNet and VNet after the fusion as well. Second, in the trade-off by fuse two phases, we increase the sensitivity by $7.01\%$ at the cost of the specificity drop by $10.81\%$, by which we value the fusion when it comes to the possible critical point of life or death for patients. As in Fig.~\ref{Fig:7264Main}, we visualize one case where our method misses the tumor prediction in the venous phase while detecting the tiny tumors successfully in the arterial phase.

From Table~\ref{Tab:FusePhaseResults}, S4C achieves the best in the sensitivity, which verifies the effectiveness of S4C framework. For the lower specific city of S4C than the classification network, we conjecture that the classification network is trained directly with a binary optimization goal and the feature map of segmentation as input can be favorable to classification network. However, the major drawback of the classification network is that it is notoriously hard to identify which regions in the original CT scans contribute to the final abnormality prediction. But, for our S4C framework, we provide radiologists with the predicted abnormal regions as a crucial cue for why we reach the decision. The convenience brought to radiologists for further diagnosis can be valued even with slightly lower specificity.

\begin{table}[t]
 \begin{center}
 \begin{tabular}{cccccc}
 \toprule
      Phase &Method &Misses &W.Calls& Sensitivity & Specificity \\
 \hline
  Arterial\&Venous &S4C (Ours) &$\mathbf{24/228}$ &$28/148$ & $\mathbf{89.47\%}$ & $81.08\%$\\
  Arterial\&Venous &3D UNet &$26/228$ &$31/148$ & $88.60\%$ & $79.05\%$\\
  Arterial\&Venous &VNet &$28/228$ &$37/148$ & $87.72\%$ & $75.00\%$\\
 \hline
  Arterial\&Venous &Classification &$24/228$ &$\mathbf{23/148}$ & $89.47\%$ & $\mathbf{84.46\%}$\\
 \bottomrule
 \end{tabular}
 \end{center}
 \caption{Performance of abnormality classification on our own dataset by considering two phases together. From left to right: the number of missed abnormal cases out of $228$ abnormal cases in total, the number of wrong calls of tumor predictions out of $148$ normal cases in total, the corresponding sensitivity and the specificity.}
 \label{Tab:FusePhaseResults}
 \vspace{-0.8cm}
\end{table}

\section{Conclusion}
In this work, we propose an overall framework to conduct the early detection of PNETs, the second common type of pancreatic cancer. We adopt the Segmentation-for-Classification strategy to make the classification result more interpretable to radiologists compared with a direct binary classification network. To quantitatively analyze our method, we voxelwisely annotate the largest PNETs CT dataset to the best of our knowledge. Furthermore, each CT case is collected in both arterial and venous phase, where the voxels of dilated pancreatic duct are annotated as well to increase the sensitivity in practice. Our approach outperforms the state-of-the-arts segmentation algorithms in terms of the DSC score and is comparable to a binary classification neural network in terms of sensitivity and specificity. In the future, we would like to integrate the classification network into the segmentation backbone, which can let these two tasks benefit from each other by a joint learning manner.

\bibliographystyle{splncs04}
\bibliography{typeinst}

\newpage
\renewcommand\thefigure{\arabic{figure}}
\setcounter{figure}{0}
\pagenumbering{gobble}
\section*{Supplementary Material}
\paragraph{\bf Global Image Caption}
The visualization illustration of predicted segmentation for ``Ours", ``3D UNet" and ``VNet" under both Venous and Arterial phases, respectively, where ``V" stands for the ``Venous" phase and ``A" stands for the ``Arterial" phase. The masked \textcolor{blue}{blue}, \textcolor{red}{red} and \textcolor{green}{green} regions denote for the normal pancreas regions, PNETs regions, and dilated pancreatic duct regions. Best viewed in color.

\vspace{-0.8cm}
\begin{figure}[h!]
	\begin{center}
		\includegraphics[width=0.8\linewidth]{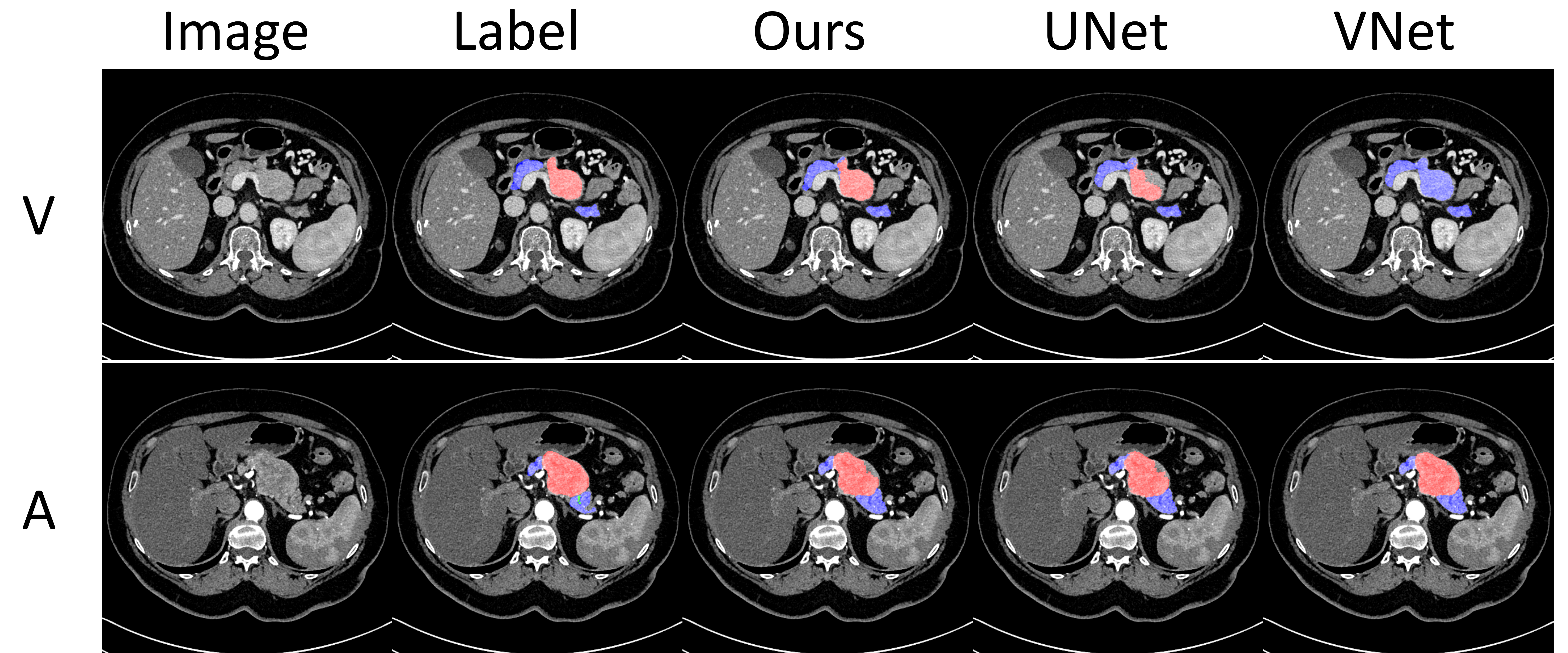} \\
	\end{center}
	\vspace{-0.8cm}
	\caption{The segmentation visualization for the case number 7011. ``Ours" method gives better performance in the venous phase while comparably good in the arterial phase. }
	\label{Fig:7011}
\end{figure}

\vspace{-1.2cm}
\begin{figure}[h!]
	\begin{center}
		\includegraphics[width=0.8\linewidth]{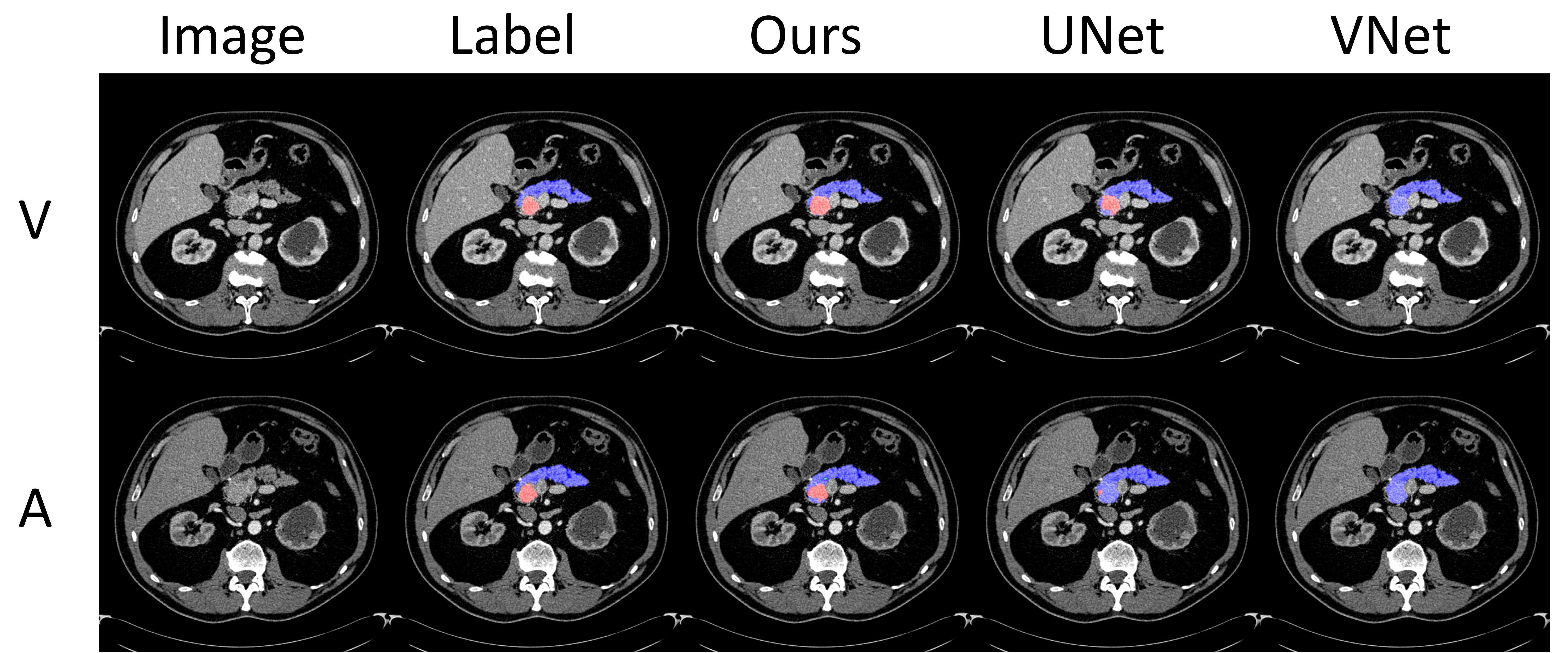} \\
	\end{center}
	\vspace{-0.8cm}
	\caption{The segmentation visualization for the case number 7328. ``Ours" method successfully detects the PNETs regions on both the venous and the arterial phase.}
	\label{Fig:7328}
\end{figure}

\vspace{-1.2cm}
\begin{figure}[h!]
	\begin{center}
		\includegraphics[width=0.8\linewidth]{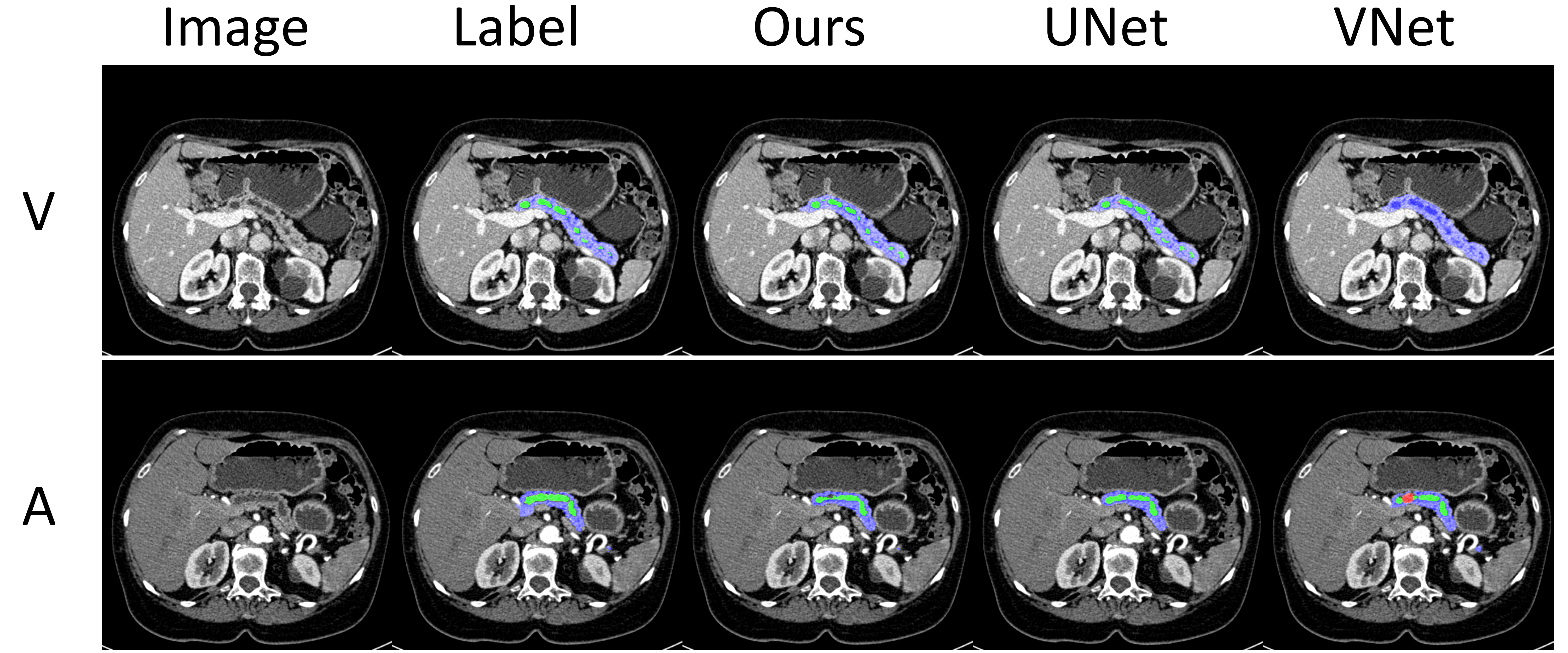} \\
	\end{center}
	\vspace{-0.8cm}
	\caption{The segmentation visualization for the case number 7349. ``Ours" method successfully detects the dilated pancreatic duct regions on both the venous and the arterial phase, which is a high risk sign for the tumors.}
	\label{Fig:7349}
\end{figure}

\vspace{-3cm}
\begin{figure}[h!]
	\begin{center}
		\includegraphics[width=0.8\linewidth]{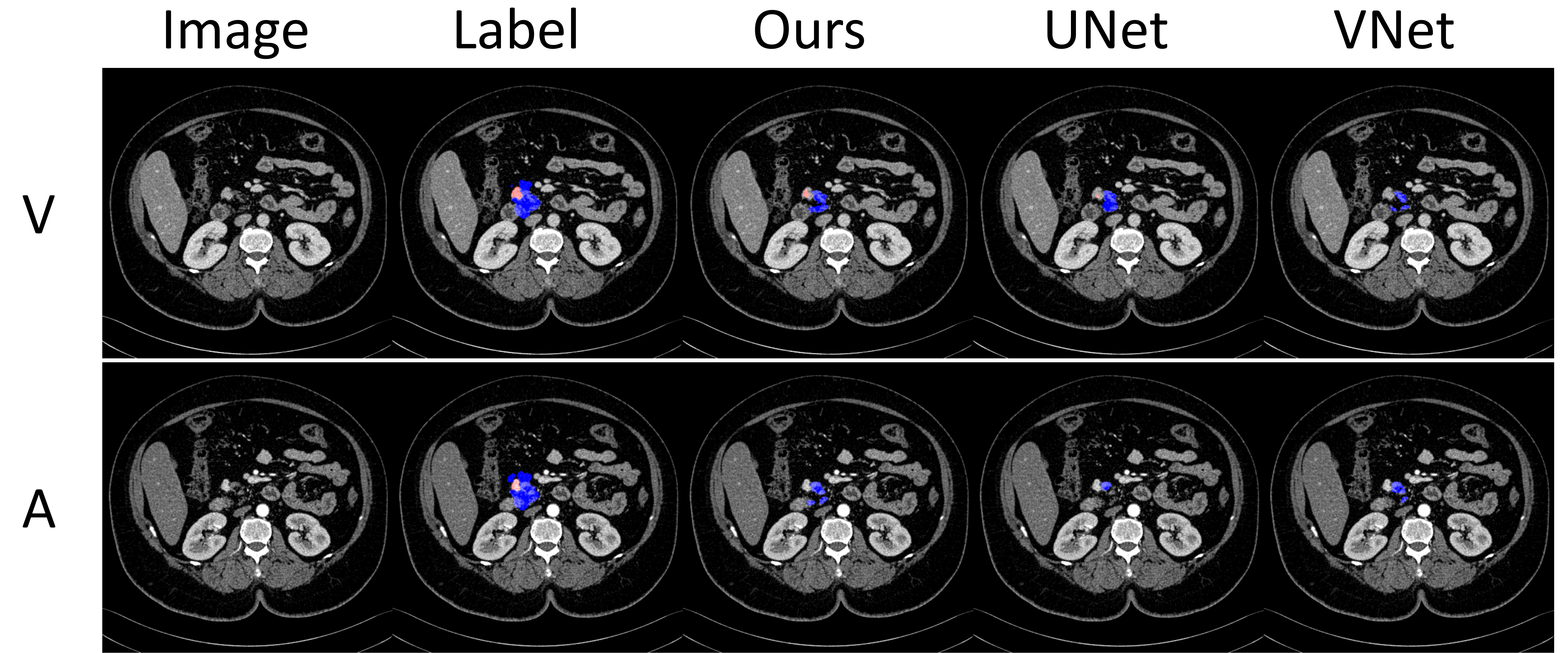} \\
	\end{center}
	\vspace{-0.8cm}
	\caption{The segmentation visualization for the case number 7086. This is a very tiny PNETs, where ``Ours" method successfully detects the PNETs regions on the venous phase while missing the detection on the arterial phase. In comparison, both ``3D UNet" and ``VNet" miss the tumor regions on both phases.}
	\label{Fig:7086}
\end{figure}

\vspace{-3cm}
\begin{figure}[h!]
	\begin{center}
		\includegraphics[width=0.8\linewidth]{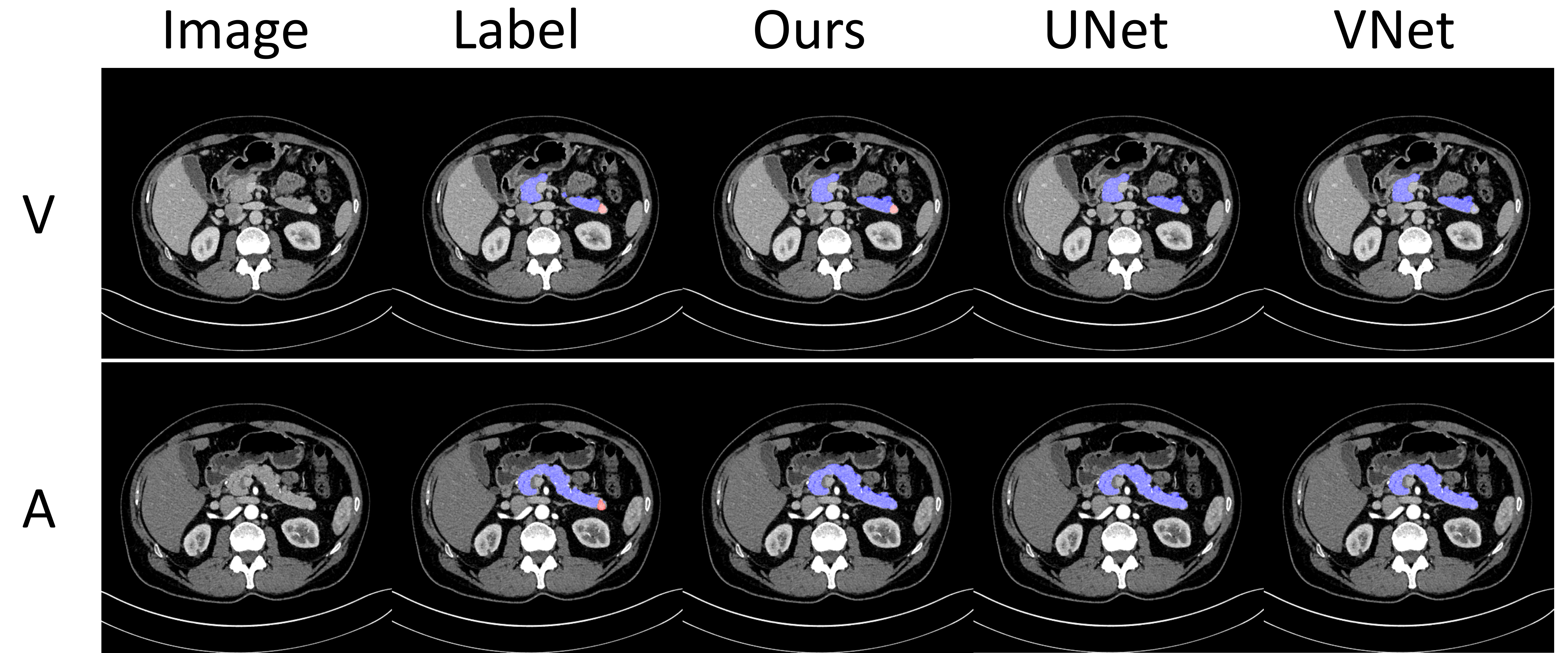} \\
	\end{center}
	\vspace{-0.8cm}
	\caption{The segmentation visualization for the case number 7207. The tiny PNETs is hanging of the pancreas tail, where only ``Ours" method successfully detects the PNETs regions on the venous phase while missing the detection on the arterial phase.}
	\label{Fig:7207}
\end{figure}

\vspace{-3cm}
\begin{figure}[h!]
	\begin{center}
		\includegraphics[width=0.8\linewidth]{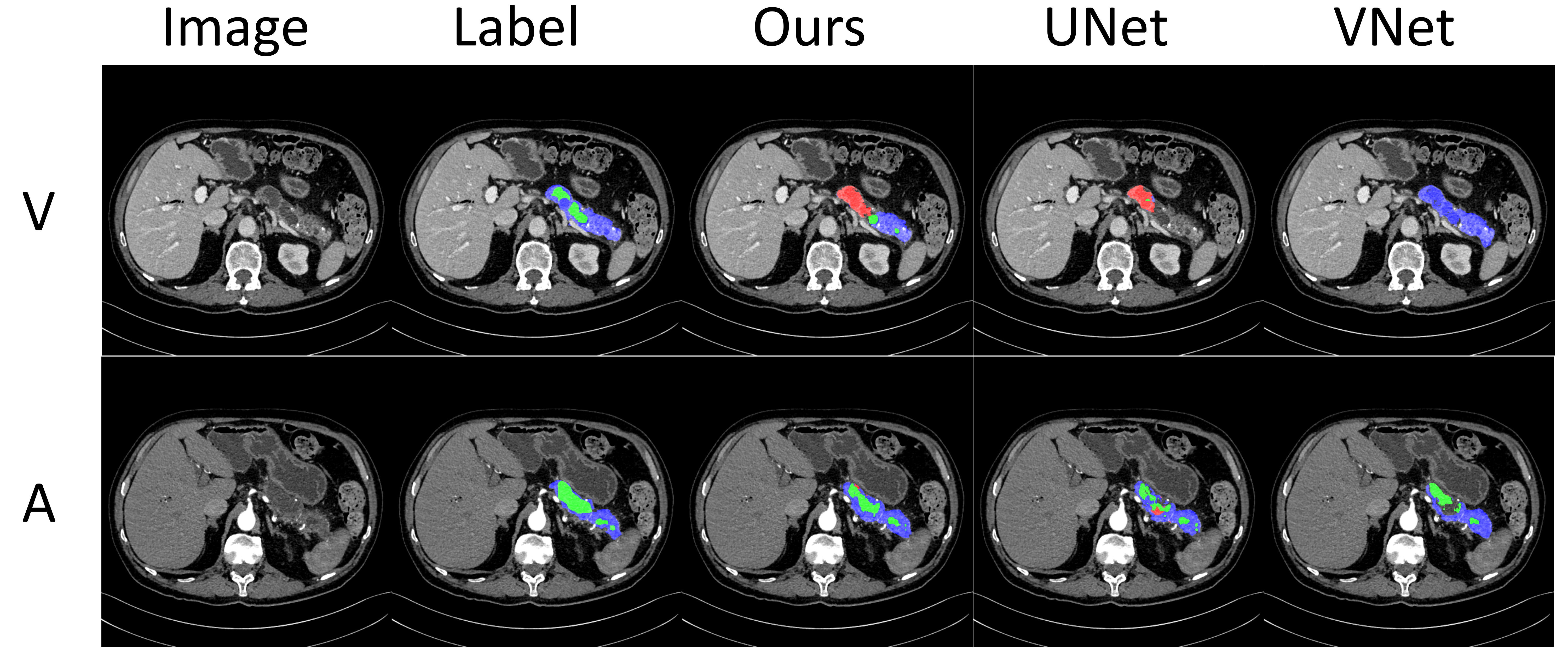} \\
	\end{center}
	\vspace{-0.8cm}
	\caption{The segmentation visualization for the case number 7179. ``Ours" method detects the dilated pancreatic duct regions fairly good on the arterial phase while mistaking the duct to be PNETs on the venous phase.}
	\label{Fig:7179}
\end{figure}

\end{document}